\documentclass[aps,prb,twocolumn,superscriptaddress,showpacs,showkeys,floatfix]{revtex4}

\usepackage{graphicx,color}

\newcommand{\jwj}[1]{\textcolor{blue}{ #1}}

\graphicspath{{figs/}}
\begin{document}
\title{Phonon Modes in Single-Walled Molybdenum Disulphide (MoS$_{2}$) Nanotubes: Lattice Dynamics Calculation and Molecular Dynamics Simulation}
\author{Jin-Wu Jiang}
    \altaffiliation{Corresponding author: jwjiang5918@hotmail.com}
    \affiliation{Shanghai Institute of Applied Mathematics and Mechanics, Shanghai Key Laboratory of Mechanics in Energy Engineering, Shanghai University, Shanghai 200072, People's Republic of China}
\author{Bing-Shen~Wang}
    \affiliation{State Key Laboratory of Semiconductor Superlattice and Microstructure and Institute of Semiconductor, Chinese Academy of Sciences, Beijing 100083, China}
\author{Timon Rabczuk}
    \altaffiliation{Corresponding author: timon.rabczuk@uni-weimar.de}
    \affiliation{Institute of Structural Mechanics, Bauhaus-University Weimar, Marienstr. 15, D-99423 Weimar, Germany}
    \affiliation{School of Civil, Environmental and Architectural Engineering, Korea University, Seoul, South Korea }

\date{\today}
\begin{abstract}
We study the phonon modes in single-walled MoS$_{2}$ nanotubes via the lattice dynamics calculation and molecular dynamics simulation. The phonon spectra for tubes of arbitrary chiralities are calculated from the dynamical matrix constructed by the combination of an empirical potential with the conserved helical quantum numbers $(\kappa, n)$. In particular, we show that the frequency ($\omega$) of the radial breathing mode is inversely proportional to the tube diameter ($d$) as $\omega=665.3/d$~{cm$^{-1}$}. The eigen vectors of the first twenty lowest-frequency phonon modes are illustrated. Based on these eigen vectors, we demonstrate that the radial breathing oscillation is disturbed by phonon modes of three-fold symmetry initially, and the tube is squashed by the modes of two-fold symmetry eventually. Our study provides fundamental knowledge for further investigations of the thermal and mechanical properties of the MoS$_{2}$ nanotubes.
\end{abstract}

\maketitle
\pagebreak

\section{Introduction}

Molybdenum Disulphide (MoS$_{2}$) is a semiconductor with a bulk bandgap above 1.2~{eV},\cite{KamKK} which can be further manipulated by changing its thickness,\cite{MakKF} or through application of mechanical strain.\cite{FengJ2012npho,LuP2012pccp} This finite bandgap is a key reason for the excitement surrounding MoS$_{2}$ as compared to graphene as graphene has a zero bandgap.\cite{NovoselovKS2005nat}  Because of its direct bandgap and also its well-known properties as a lubricant, MoS$_{2}$ has attracted considerable attention in recent years.\cite{WangQH2012nn,ChhowallaM} For example, Radisavljevic et al.\cite{RadisavljevicB2011nn} demonstrated the application of single-layered MoS$_{2}$ (SLMoS$_{2}$) as a good transistor. The strain and the electronic noise effects were found to be important for the SLMoS$_{2}$ transistor.\cite{ConleyHJ,SangwanVK,Ghorbani-AslM,CheiwchanchamnangijT}

Besides electronic properties, there has been increasing interest in the thermal and mechanical properties of MoS$_{2}$. Several recent works have addressed the thermal transport properties of SLMoS$_{2}$ in both ballistic and diffusive transport regimes.\cite{HuangW,JiangJW2013mos2,VarshneyV,JiangJW2013sw} The mechanical behavior of SLMoS$_{2}$ has been investigated experimentally.\cite{BertolazziS,CooperRC2013prb1,CooperRC2013prb2} For the theoretical part, we have examined the size and edge effects on the Young's modulus of the SLMoS$_{2}$ based on the Stillinger-Weber (SW) potential.\cite{JiangJW2013sw} Quite recently, we derived an analytic formula for the elastic bending modulus of the SLMoS$_{2}$, where the importance of the finite thickness effect was revealed.\cite{JiangJW2013bend}

A fundamental property related to the thermal and mechanical behaviors is the lattice dynamics property of MoS$_{2}$, i.e the phonon spectrum or phonon modes. The thermal conductivity in the semiconductor MoS$_{2}$ is dominated by the lattice thermal transport, which is contributed by the phonon mode. Different mechanical processes are governed by the corresponding phonon modes. For instance, the nanomechanical resonant oscillation takes advantage of the bending mode (flexure mode). Recently, there have been increasing experimental and theoretical efforts on the lattice dynamics properties of the single-layer or few-layer MoS$_{2}$ nanosheets.\cite{WakabayashiN,JimenezSS,LeeC,SanchezAM,ZengHL,ZhaoYY,RiceC,ZhangXprb2013,LanzilloN2013} However, up to date, the lattice dynamics of the single-walled MoS$_{2}$ nanotube (SWMoS$_{2}$NT) was investigated only by few works, which show different lattice properties in the tube structure from the planar sheet structure, because of the special line group in the nanotube.\cite{DobardzicE,DamnjanovicM2008mmp}

In this paper, we perform the lattice dynamics study and the molecular dynamics simulation for the lattice properties of the SWMoS$_{2}$NT with arbitrary chiralities. The phonon spectrum is calculated from the dynamics matrix with the force constant matrix obtained from the SW potential. The helical quantum numbers are used to denote the phonon mode. The eigen vector of the first twenty lowest-frequency phonon modes are presented. Furthermore, we show that the radial breathing mechanical oscillation is initially disturbed by the three-fold modes and eventually destroyed by the two-fold modes.

\begin{figure}[htpb]
  \begin{center}
    \scalebox{0.9}[0.9]{\includegraphics[width=8cm]{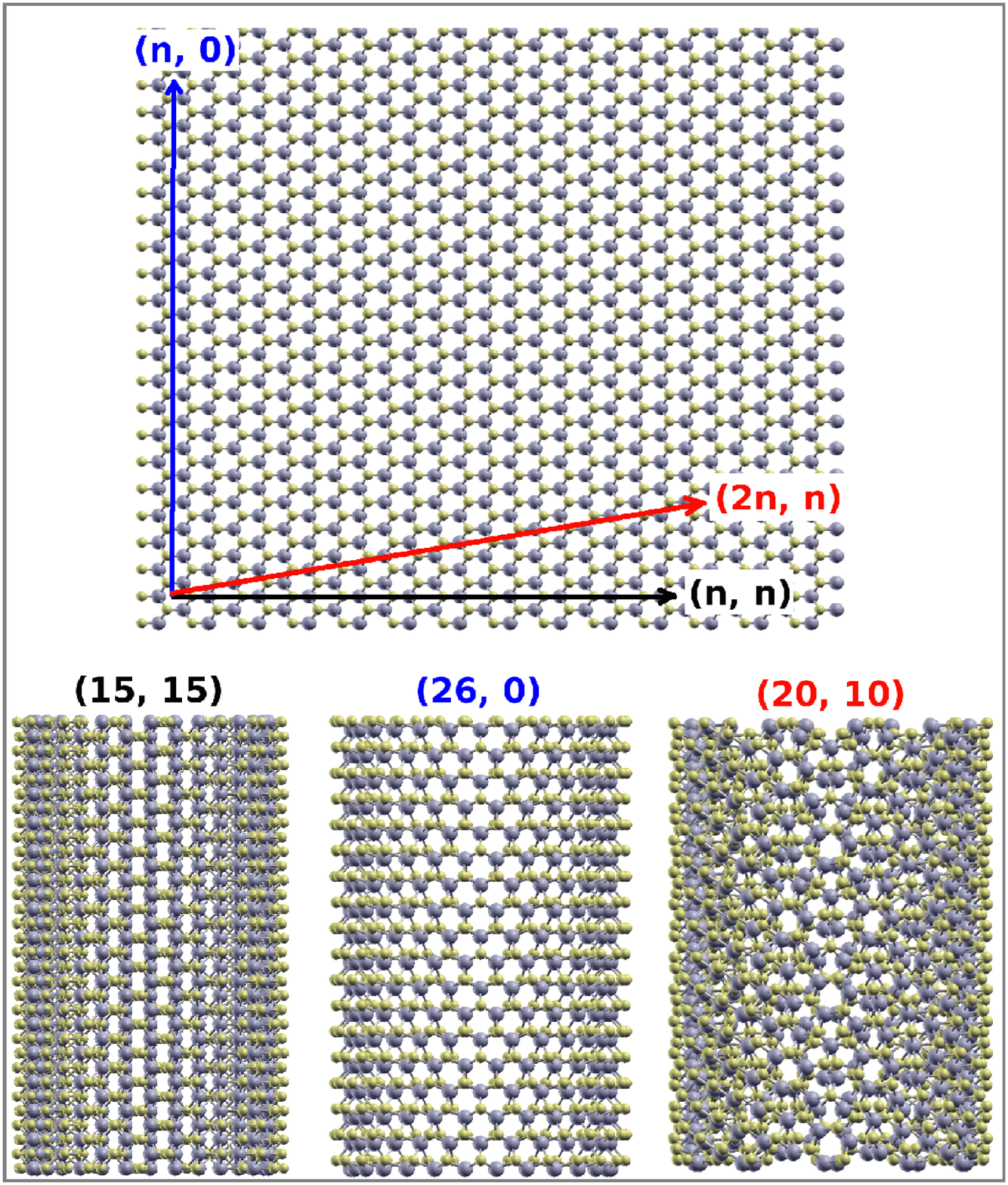}}
  \end{center}
  \caption{(Color online) The configuration of MoS$_{2}$. Mo atoms are represented by big balls (gray online). S atoms are presented by small balls (yellow online). Above panel: top view of the two-dimensional MoS$_{2}$ honeycomb lattice. Three lattice directions are depicted by arrows, $\vec{R}_{\rm arm}=n_{1}\vec{a}_{1}+n_{1}\vec{a}_{2}$, $\vec{R}_{\rm zig}=n_{1}\vec{a}_{1}$, and $\vec{R}_{\rm chiral}=2n_{2}\vec{a}_{1}+n_{2}\vec{a}_{2}$. Bottom panels: three SWMoS$_{2}$NTs are obtained by rolling up the above MoS$_{2}$ sheet onto a cylindrical surface with the three corresponding lattice vectors as the circumference.}
  \label{fig_cfg}
\end{figure}

\begin{figure*}[htpb]
  \begin{center}
    \scalebox{0.8}[0.8]{\includegraphics[width=\textwidth]{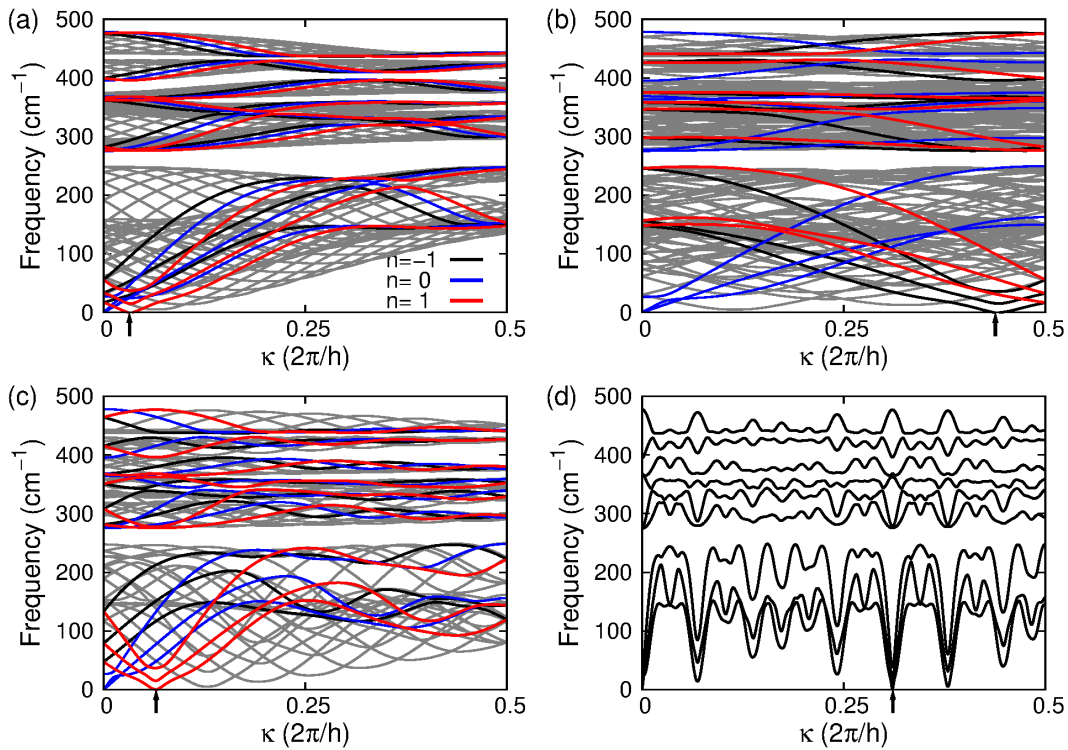}}
  \end{center}
  \caption{(Color online) Phonon spectra in SWMoS$_{2}$NTs of different chiralities. (a) Armchair (15, 15). (b) Zigzag (26, 0). (c) Chiral (20, 10). (d) Chiral (16, 13). The x axis is the helical quantum number $\kappa$ corresponding to the screw symmetric operation. The number $n$ is the other quantum number corresponding to the pure rotational symmetric operation. For panels (a), (b), and (c), $n=-1, 0, 1$ are shown by black, blue, and red lines, respectively. \jwj{Other general curves are displayed by gray lines.} In panel (d), only a single value for $n$, i.e $n=0$. The arrow in each figure depicts the position of one of the TA (flexure) mode. The other flexure mode locates at the $\kappa$ with the same magnitude but negative sign. The spectrum is overall symmetric about $\kappa=0$, so only curves in $\kappa\in[0, 0.5]$ are displayed.}
  \label{fig_dispersion}
\end{figure*}

\begin{figure*}[htpb]
  \begin{center}
    \scalebox{0.8}[0.8]{\includegraphics[width=\textwidth]{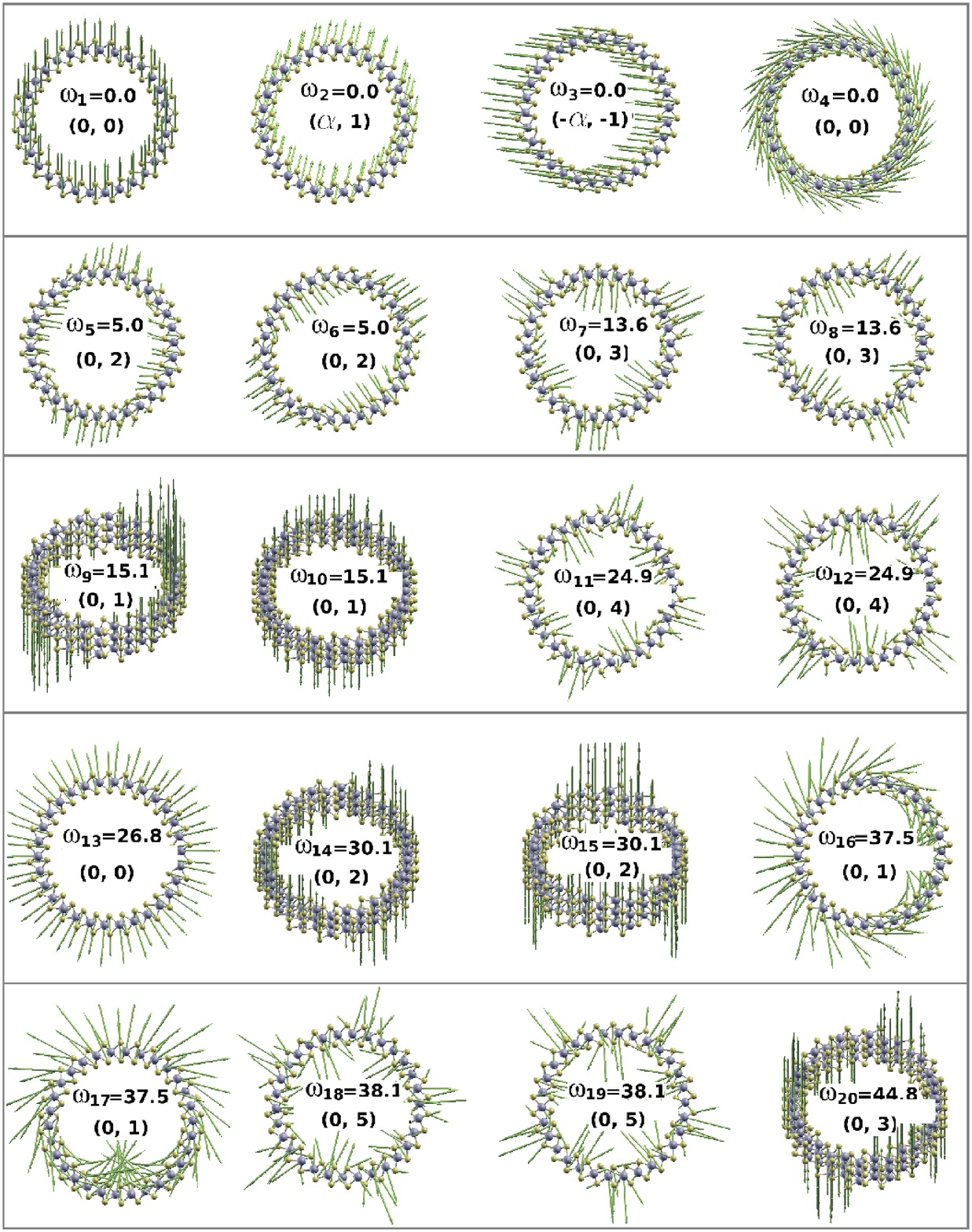}}
  \end{center}
  \caption{(Color online) The first twenty lowest-frequency phonon modes in SWMoS$_{2}$NT (15, 15). Frequencies are given in the unit of cm$^{-1}$. Numbers in the parenthesis are the helical quantum numbers $(\kappa n)$ for each mode. The arrow attached to each atom represents the component of this atom in the eigen vector of the phonon mode. The view direction is along the axial direction.}
  \label{fig_u_arm}
\end{figure*}

\begin{figure*}[htpb]
  \begin{center}
    \scalebox{0.8}[0.8]{\includegraphics[width=\textwidth]{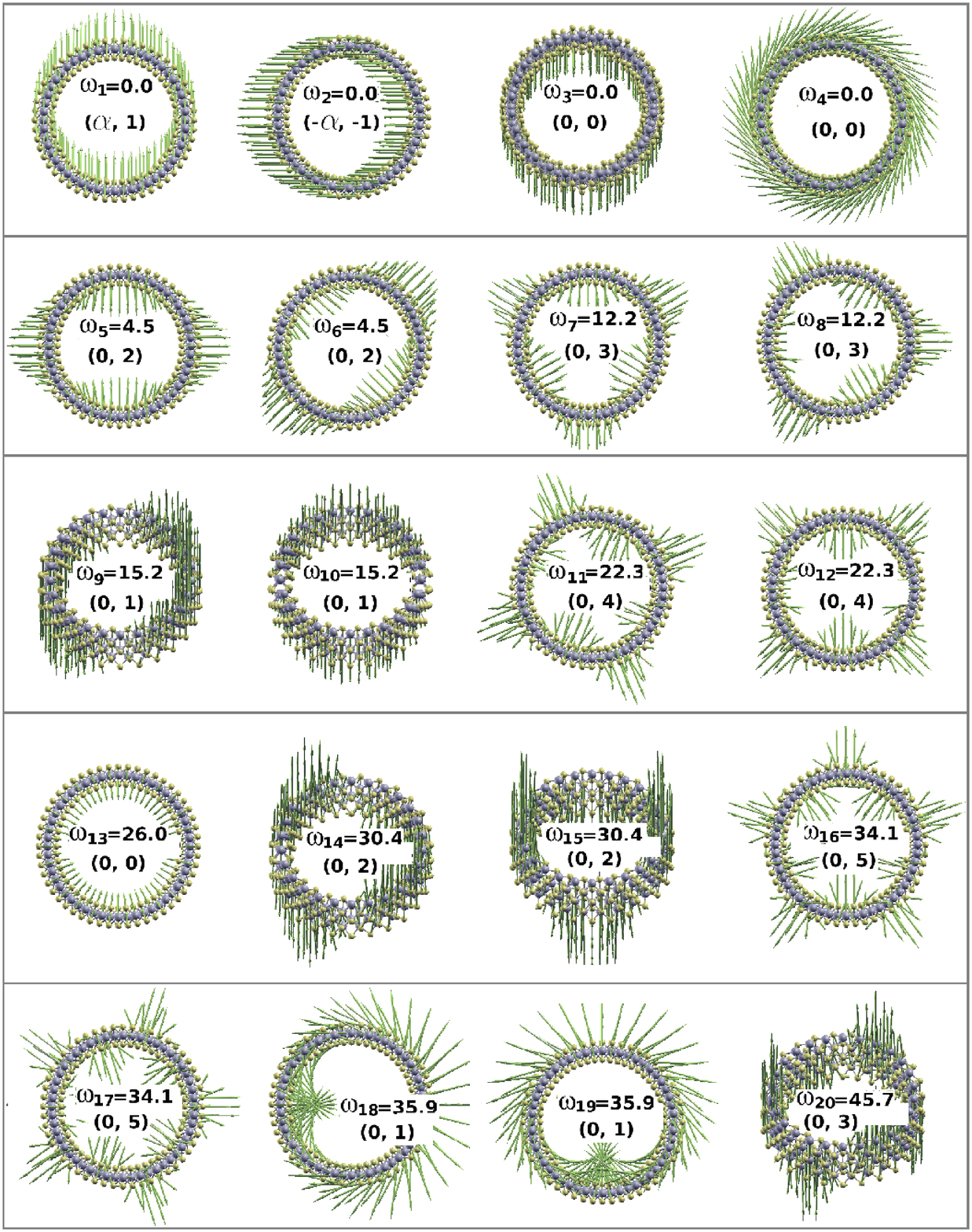}}
  \end{center}
  \caption{(Color online) The first twenty lowest-frequency phonon modes in SWMoS$_{2}$NT (26, 0). Frequencies are given in the unit of cm$^{-1}$. Numbers in the parenthesis are the helical quantum numbers $(\kappa n)$ for each mode. The arrow attached to each atom represents the component of this atom in the eigen vector of the phonon mode. The view direction is along the axial direction.}
  \label{fig_u_zig}
\end{figure*}

\begin{figure*}[htpb]
  \begin{center}
    \scalebox{0.8}[0.8]{\includegraphics[width=\textwidth]{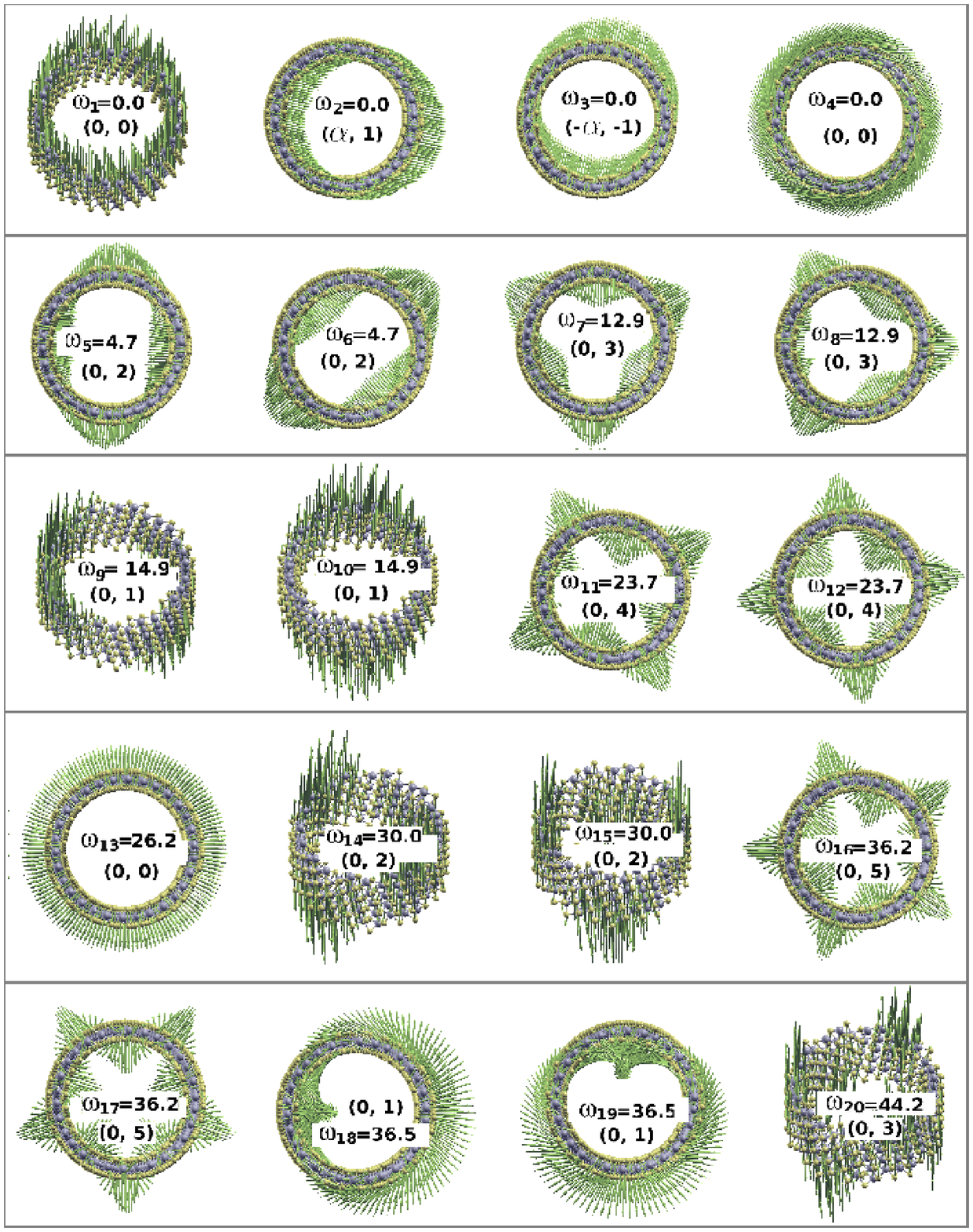}}
  \end{center}
  \caption{(Color online) The first twenty lowest-frequency phonon modes in SWMoS$_{2}$NT (20, 10). Frequencies are given in the unit of cm$^{-1}$. Numbers in the parenthesis are the helical quantum numbers $(\kappa n)$ for each mode. The arrow attached to each atom represents the component of this atom in the eigen vector of the phonon mode. The view direction is along the axial direction.}
  \label{fig_u_chiral}
\end{figure*}

\section{Structure and Calculation Details}
\jwj{The top view of the SLMoS$_{2}$ is shown in Fig.~\ref{fig_cfg} (top panel). The three arrows indicate three representative lattice directions: armchair lattice direction $(n_{1}, n_{1})$, zigzag lattice direction $(n_{1}, 0)$, and a chiral lattice direction $(2n_{2}, n_{2})$. The lattice constant is $|\vec{a}_{1}|=2.14$~{\AA}. The SLMoS$_{2}$ can be rolled up into tubal structures shown in the bottom of Fig.~\ref{fig_cfg}.}

\jwj{Besides the two primitive vectors $\vec{a}_{1}$ and $\vec{a}_{2}$, the lattice structure can be described by any other pair of equivalent primitive lattice vectors, which surround the same area $|\vec{a}_{1}\times\vec{a}_{2}|$. A particular pair of such equivalent primitive lattice vectors are $\vec{R}_{H}$ and $\frac{\vec{R}}{N}$,\cite{WhiteCT} which have been used for SWMoS$_{2}$NT.\cite{MilosevicI} The helical lattice vector $\vec{H}=p_{1}\vec{a}_{1}+p_{2}\vec{a}_{2}$ describing the screw operation, with $n_{1}p_{2}-n_{2}p_{1}=N$. $N$ is the greatest common divisor of $n_{1}$ and $n_{2}$. A big translational cell along the axial direction can be produced by a proper application of the screw and rotational operations.\cite{WhiteCT} A unit cell in the $(n_{1}, n_{2})$ tubule can be notated in two equivalent ways, either by the primitive vector $(\vec{H}, \frac{\vec{R}}{N})$, or by $(\vec{a}_1, \vec{a}_2)$ as}
\begin{eqnarray}
      \vec{r}_{m,l}=m\vec{H}+l\frac{\vec{R}}{N}, \hspace{0.25cm}
      \mbox{or} \hspace{0.25cm}
      \vec{r}_{q_{1},q_{2}}=q_{1}\vec{a}_{1}+q_{2}\vec{a}_{2},
     \label{eq_rvec}
\end{eqnarray}
in which $(q_{1}, q_{2})$ and $(m l)$ are related to each other as:
\begin{eqnarray}
     m=(n_{1}q_{2}-n_{2}q_{1})/N, \hspace{0.25cm}
     l=q_{1}p_{2}-q_{2}p_{1}\;.
     \label{eq_ml}
\end{eqnarray}
$\vec{b}_{H}$ and $\vec{b}_{R}$ are reciprocal unit vectors corresponding to $(\vec{H}, \frac{\vec{R}}{N})$. Any wave vector in the reciprocal space can be written as 
\begin{eqnarray}
\vec{K}=\frac{\kappa}{2\pi}\vec{b}_{H}+\frac{n}{N}\vec{b}_{R}\;.
\end{eqnarray}
$(\kappa n)$ are the two helical quantum numbers. In the first Brillouin zone, $\kappa\in(-\frac{1}{2}, \frac{1}{2}]$ and $n$ is an integer in $(-\frac{N}{2}, \frac{N}{2}]$. \jwj{A comprehensive description for the relationship between different sets of symmetric notations can be found in Ref.~\onlinecite{DobardzicE2003prb} and in the book chapter by Tang, Wang, and Su.\cite{TangH2011book}}

\section{Phonon Dispersion of SWMoS$_{2}$NT}
All screw and rotational symmetry operations in the SWMoS$_{2}$NT form the space group of this tubal system, which is named line group.\cite{VujicicM,BozovicIB1978jpamg,BozovicIB1981jpamg,MilosevicI}  According to the irreducible representation of the line group, we have the generalized Bloch theory for the vibration displacement of each atom $(mls)$ in the phonon mode $(\kappa n \tau)$,\cite{MiloevicI1993prb,PopovVN1999prb,DobardzicE2003prb,JiangJW2006,DakicB2009jpamt}
\begin{eqnarray*}
\vec{u}(mls) & = & \frac{1}{\sqrt{MN}}\sum_{\kappa n \tau}e^{i\left(\kappa m+\frac{2\pi}{N}nl\right)}R\left(ml\right)\vec{\xi}^{(\tau)}(\kappa n|00s)\hat{Q}_{\kappa n}^{\tau}
\label{eq_bloch}
\end{eqnarray*}
where $\tau$ is the branch index. $M$ is the total number of the screw symmetry operation. $N$ is the greatest common divisor of $n_{1}$ and $n_{2}$. It is the number of pure rotational symmetry operations. $M\times N$ gives the total number of unit cells in the tube. $s=1,2,3$ corresponds to the three atoms S, Mo, and S in the unit cell. $R\left(ml\right)$ is a rotation matrix for rotation around the axial direction for angle $\phi_{ml}  =  m\alpha+l\frac{2\pi}{N}$. $\vec{\xi}^{(\tau)}(\kappa n|00s)$ is the vibration displacement for atom $(00s)$ in the phonon mode denoted by $\left(\kappa n \tau\right)$. $\hat{Q}_{\kappa n}^{\tau}$ is the helical quantum state of the phonon mode $\left(\kappa n \tau\right)$.

Applying the generalized Bloch theory into the phonon dynamical equations, we can obtain the dynamics matrix of the SWMoS$_{2}$NT as,
\begin{eqnarray*}
D_{ss'}\left(\kappa n\right) & = & \frac{1}{\sqrt{m_{s}m_{s'}}}\sum_{(mls')}\Phi(00s|mls')R\left(ml\right)e^{i\left(\kappa m+\frac{2\pi}{N}nl\right)},
\label{eq_D}
\end{eqnarray*}
where $m_{s}$ is the mass of atom $s$. For MoS2, the dynamical matrix is a $9\times9$ matrix, since there are three atoms in each unit cell. The diagonalization of the dynamical matrix results in nine phonon modes denoted by the helical index $\left(\kappa n \tau\right)$, with $\tau=1,2,3,...,9$.

$\Phi(00s|mls')$ is the force constant matrix. In our calculation, this force constant matrix is obtained using finite differential method based on a newly developed SW potential. The SW potential is developed for SLMoS$_{2}$, including two-body and three-body interactions. We apply this potential to describe the atomic interaction within the SWMoS$_{2}$NT. All SW potential parameters can be found in our previous work.\cite{JiangJW2013sw}

Fig.~\ref{fig_dispersion} shows the phonon spectrum in four SWMoS$_{2}$NTs with almost the same diameter of 18.0~{\AA}. The spectrum is symmetric about $\kappa=0$ in all tubes, so only curves with $\kappa\in[0, 0.5]$ are displayed. Panel (a) is the phonon spectrum for armchair tube (15, 15). There are nine curves for each helical quantum number $n$, corresponding to the nine degrees of freedom in each unit cell. There are four acoustic branches. The longitudinal acoustic (LA) and twisting (TW) phonon branches have the same helical quantum number $n=0$. The other two acoustic branches are the transverse acoustic (TA) phonons with $n=\pm 1$. One of the TA mode locates at $\kappa=\alpha$ and $n=1$, which is depicted by a solid arrow on the $\kappa$ axis. The other TA mode locates at $\kappa=-\alpha$ and $n=-1$. The two TA modes are also called flexure modes, and they have a parabolic spectrum due to the quasi-one-dimensional nature of the nanotubes. \jwj{The TW mode is a characteristic for the cylindrical geometry of the SWMoS$_{2}$NT. The interaction is invariant during the rotation of the SWMoS$_{2}$NT around its axis, leading to the particular (TW) acoustic mode in the SWMoS$_{2}$NT.} Panel (b) shows the phonon spectrum of the zigzag tube (26, 0). Panels (c) and (d) show the phonon spectrum for chiral tubes. In tube (16, 13), the quantum number $n$ has only one value, i.e $n=0$, so there is only nine curves in this figure.

One significant feature of Fig.~\ref{fig_dispersion} is the combination of an empirical many-body potential (SW) with the helical quantum numbers $(\kappa n)$. In previous studies, the helical quantum numbers are used only in dynamical matrix constructed from force field models. \jwj{For dynamical matrix constructed from the empirical potential, the phonon spectrum is denoted by the linear wave vector corresponding to the big translation symmetry and the angular quantum number corresponding to pure rotational symmetries. These quantum numbers are related to the helical quantum numbers.\cite{DobardzicE2003prb}} The empirical potential can give more comprehensive information for the phonon spectrum. In particular, it fulfills the rigid rotational symmetry, leading to the zero frequency of the TW mode. Furthermore, the helical quantum numbers correspond to the actual symmetric operations (screw and pure rotational symmetric operations) in the SWMoS$_{2}$NT. Hence they are the good quantum numbers for the phonon modes in the system.\cite{DobardzicE2003prb} The phase `good' means that these quantum numbers are conserved during the phonon-assisted physical process. For instance, it has been shown that these helical quantum numbers give a natural (the simplest) selection rule for the phonon-phonon scattering process in the lattice thermal transport of the single-walled boron nitride nanotube.\cite{JiangJW2011bntube} Here, we have successfully combined the helical quantum numbers with empirical potential; thus take advantage of both parts. Hence, Fig.~\ref{fig_dispersion} provides a `good' recipe for further investigations of the phonon-assisted physical processes in SWMoS$_{2}$NTs.

We now pay attention to the low-frequency modes in SWMoS$_{2}$NTs, because these modes are easier to be excited in practice. Fig.~\ref{fig_u_arm} shows the first twenty lowest-frequency modes in the armchair tube (15, 15). Three big translational cells are used in this calculation. The periodic boundary condition is applied in the axial direction. These modes locate in the $\Gamma$ point in the Brillouin zone; i.e, the linear quantum number is zero. The linear quantum number is the wave vector corresponding to the big translational operation. Numbers in the figure are the frequency of each mode in the unit of cm$^{-1}$. The figure is plotted with XCRYSDEN.\cite{xcrysden} The arrow attached to each atom represents the component of this atom in the eigen vector of the phonon mode.

\begin{figure}[htpb]
  \begin{center}
    \scalebox{1.0}[1.0]{\includegraphics[width=8cm]{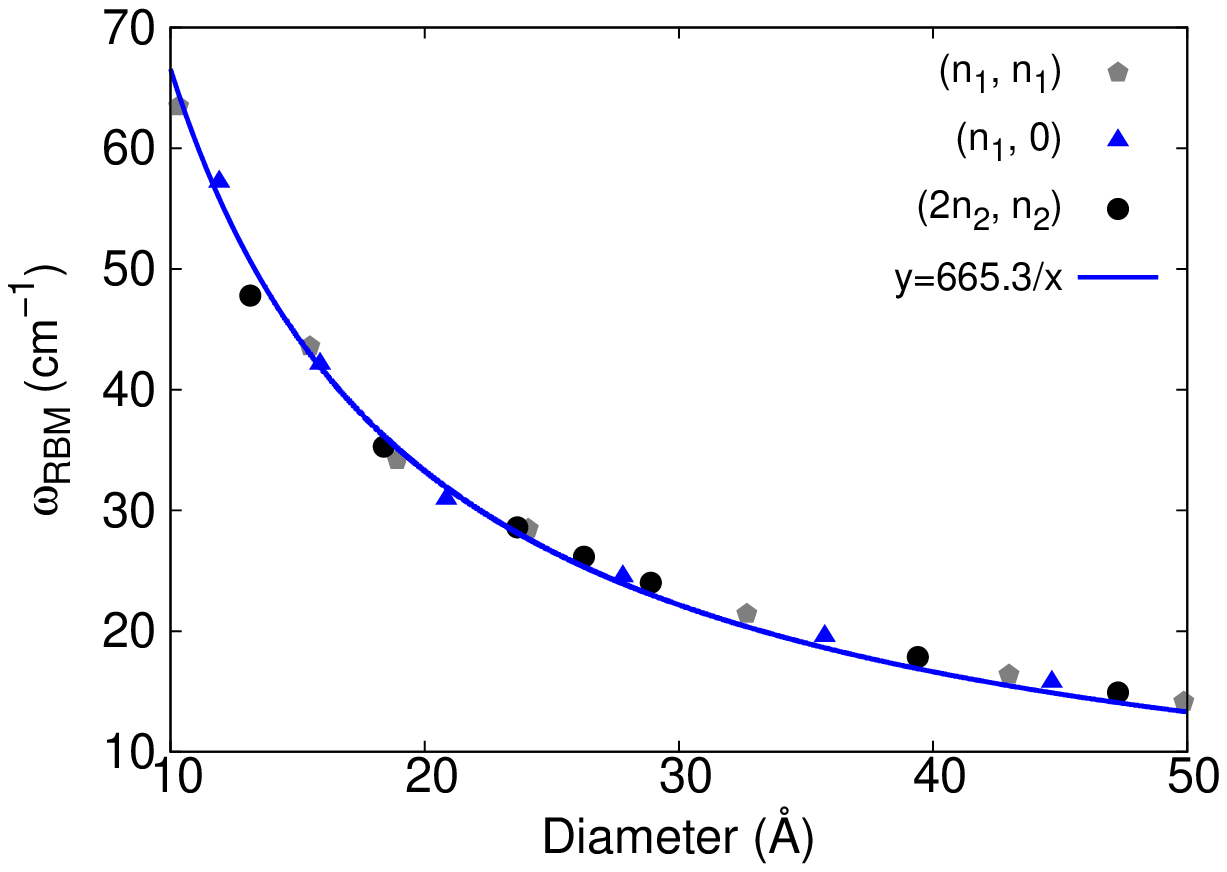}}
  \end{center}
  \caption{(Color online) The diameter dependence for the frequency of the RBM in armchair, zigzag, and chiral tubes. The solid line (blue online) is the fitting function, $y=665.3/x$, for all data.}
  \label{fig_frequency}
\end{figure}

\begin{figure}[htpb]
  \begin{center}
    \scalebox{0.9}[0.9]{\includegraphics[width=8cm]{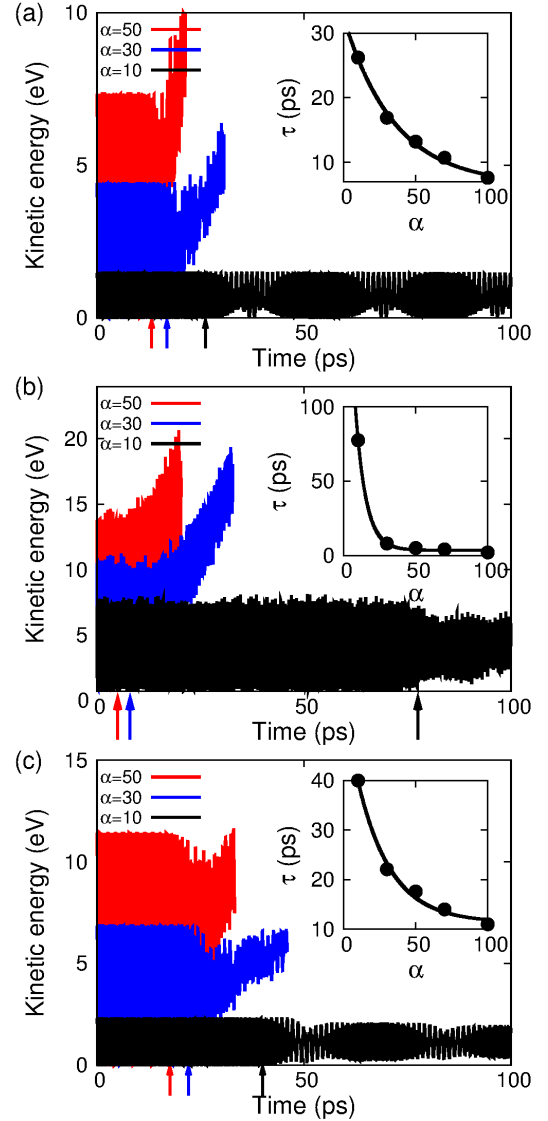}}
  \end{center}
  \caption{(Color online) The total kinetic energy time history during the radial breathing oscillation in SWMoS$_{2}$NTs of different chiralities. Tubes oscillate like a breather. (a) Armchair (15, 15). (b) Zigzag (26, 0). (c) Chiral (20, 10). Arrows depict the life time, at which the radial breathing oscillation starts to be disturbed. The inset in each panel shows the life time versus actuation parameter $\alpha$. The solid line is the exponential fitting to the calculated data.}
  \label{fig_energy}
\end{figure}

\begin{figure*}[htpb]
  \begin{center}
    \scalebox{0.8}[0.8]{\includegraphics[width=\textwidth]{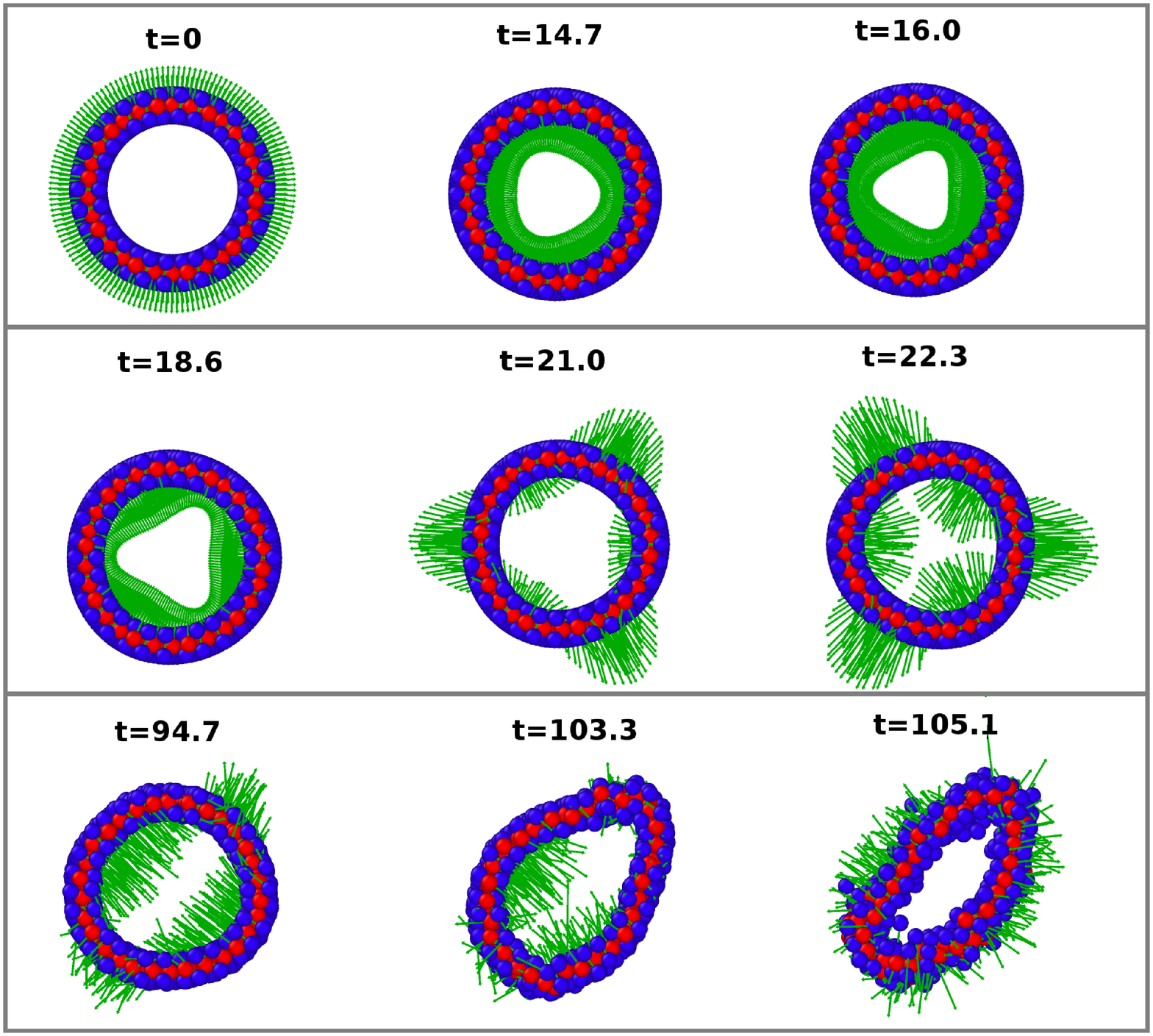}}
  \end{center}
  \caption{(Color online) The evolution of the configuration for SWMoS$_{2}$NT (20, 10) during the radial breathing oscillation with actuation parameter $\alpha=50.0$. The arrow attached to each atom represents the velocity of the atom. At $t=0$~{ps}, the oscillation is actuated by adding an initial velocity distribution, which follows the eigen vector of the RBM (13rd mode) in Fig.~\ref{fig_u_chiral}. Configurations at $t=14.7$ and 16.0~{ps} show that the radial breathing oscillation is gradually disturbed by the three-fold modes (7th and 8th modes) in Fig.~\ref{fig_u_chiral}. The radial breathing oscillation is completely replaced by the three-fold mode after $t=18.6$~{ps}. The three-fold oscillation is clearly demonstrated by the two configurations at $t=21.0$ and 22.3~{ps}, which exactly follow the eigen vector of the two three-fold modes. After $t=94.7$~{ps}, the two-fold modes (5th and 6th modes) in Fig.~\ref{fig_u_chiral} start to be excited. The tube starts to be squashed by these two-fold modes after $t=105.1$~{ps}.}
  \label{fig_cfg_md}
\end{figure*}

The first four modes are the acoustic phonon modes with zero frequency. Their helical quantum numbers are $(\kappa n) = (0, 0)$, $(\alpha, 1)$, $(-\alpha, -1)$, and $(0, 0)$. In particular, the fourth mode is the TW mode, resulting from the particular hollow cylindrical structure of the SWMoS$_{2}$NT. These four acoustic modes have important contribution to the thermal transport of SWMoS$_{2}$NTs. The fifth and sixth modes have the two-fold symmetry, which are denoted by $(\kappa n)=(0,2)$. From their peculiar vibration morphology and their pretty low frequencies, it can be easily imagined that the vibration of these modes is likely to squash the tube configuration, leading to possible instability of the SWMoS$_{2}$NT. The seventh and eighth modes have a three-fold symmetry, which are denoted by $(\kappa n)=(0,3)$. The thirteenth mode is the radial breathing mode (RBM), which is denoted by $(\kappa n)=(0,0)$. The frequency of the RBM is sensitive to the tube diameter, due to its vibration morphology and this mode is Raman active, so its frequency is usually used to estimate the tube diameter in the experiment. Figs.~\ref{fig_u_zig} and ~\ref{fig_u_chiral} show the first twenty lowest-frequency modes in tubes (26, 0) and (20, 10), respectively. In particular, the RBM has close frequency among these three tubes of quite different chiralities, since the chirality only takes effect on the order of\cite{JiangJW2006} $1/r^{3}$.

\section{RBM mode}
In the above, we have studied the full phonon spectrum for SWMoS$_{2}$NTs. The rest of this paper is devoted to the discussion of the RBM, since this mode plays an important role in the tubal structure. We will study the diameter dependence and nonlinear properties of the RBM. Fig.~\ref{fig_frequency} shows the frequency of the RBM in armchair, zigzag, and chiral tubes, i.e $(n_{1},n_{1})$, $(n_{1},0)$, and $(2n_{2},n_{2})$. \jwj{The periodic boundary condition is applied in the axial direction, since the tube length is longer than the interaction range. There are three big translational cells along the axial direction for the armchair $(n_{1},n_{1})$ and zigzag $(n_{1},0)$ tubes. One big translational cell is considered for the chiral tube $(2n_{2},n_{2})$,  which has a length of 14.3~{\AA}. This length is much larger than the interaction range of 4.27~{\AA} for the SW potential used in the simulation; thus is large enough to avoid possible edge effects.}

The frequency of all tubes can be fitted to function $\omega=665.3/d$~{cm$^{-1}$}, where $d$ is the diameter. The frequency of the RBM in SWMoS$_{2}$NTs is much lower than that in the single-walled carbon nanotube with the same diameter. The coefficient here 665.3 is about one third of the value of 2295.6 in the single-walled carbon nanotube.\cite{PopovVN1999prb} It is because the frequency of the RBM is related to the tensile mechanical properties,\cite{RaravikarNR} and the mechanical strength is weaker in the SWMoS$_{2}$NT.\cite{JiangJW2013sw} Furthermore, Mo and S atoms are heavier than the C atom in carbon nanotubes. Fig.~\ref{fig_frequency} can be useful in the estimation of the diameter for SWMoS$_{2}$NTs in the experiment.

We further investigate the nonlinear effect on the RBM. The molecular dynamics is performed to simulate the radial breathing oscillation of SWMoS$_{2}$NTs (15, 15), (26, 0), and (20, 10). The periodic boundary condition is applied in the axial direction. The tube is optimized to the energy minimum configuration. At the optimized configuration, the initial velocity distributions are set according to the vibration morphology of the RBM. The total kinetic energy corresponding to this initial velocity distribution is $\Delta E =\alpha NE_{0}$, where $\alpha$ is the energy actuation parameter.\cite{JiangJW2012jap} $N$ is the total atom number. $E_{0}=0.54$~{meV} is a small piece of energy quantum for a single atom. The tube is allowed to oscillate within the NVE ensemble with the initial velocity distribution. The total kinetic energy and the potential energy exchange between each other during this radial breathing oscillation.

In the initial oscillation stage, there is only a single oscillation mode, i.e the radial breathing oscillation. If the oscillation is in the linear regime, i.e it has a small amplitude, this radial breathing oscillation can be preserved for a very long time and other vibration modes are seldom excited. However, if the oscillation is in the nonlinear regime, i.e with larger actuation parameter $\alpha$, then the other oscillation modes will be excited quickly. As a result, the radial breathing oscillation decays quickly, owning to the nonlinear induced phonon-phonon scattering between the RBM and the other excited modes.

Fig.~\ref{fig_energy} shows the kinetic energy time history during the radial breathing oscillation of the three tubes. To study the nonlinear effect, we have shown only simulations with large actuation parameters $\alpha=10$, 30, and 50 in the figure. The radial breathing oscillation will not decay if very small actuation parameter is used, eg. $\alpha=1.0$, which will be similar as the graphene nanomechanical resonator.\cite{JiangJW2012nanotechnology}  Fig.~\ref{fig_energy} shows that in the initial stage the total kinetic energy oscillates between zero and a maximum value, which reflects the good resonant oscillation of the SWMoS$_{2}$NT. After some time, the radial breathing oscillation starts to be disturbed by some other vibration modes, which are excited by the nonlinear effect relating to the large actuation energy. Arrows in the figure depict the life time, at which the radial breathing oscillation starts to be disturbed. The inset in each panel shows the life time versus actuation parameter $\alpha$. The inset shows that the life time decays exponentially with increasing actuation parameter $\alpha$. The solid line is the exponential fitting to the calculated data. This set of simulations display that it is important to use small actuation parameter for the radial breathing oscillation, if this oscillation is used as the nanomechanical resonator.\cite{RaravikarNR}

In all of the three tubes, for very large actuation parameters such as $\alpha=30$ and 50, the kinetic energy increases sharply after the radial breathing oscillation is disturbed. It indicates that these tubes undergo a structure transition, which releases lots of potential energy. This potential energy is converted into the kinetic energy in the system. Fig.~\ref{fig_cfg_md} illustrates the structure evolution during the radial breathing oscillation in the SWMoS$_{2}$NT (20, 10) with actuation parameter $\alpha=50.0$. These snap-shots are produced by OVITO.\cite{ovito} Similar phenomena have been observed in the molecular dynamics simulation of tubes (15, 15) and (26, 0). The arrow attached to each atom represents the velocity of the atom at that moment. At $t=0$~{ps}, the radial breathing oscillation is actuated by adding an initial velocity distribution, which follows the eigen vector of the RBM (13rd mode) in Fig.~\ref{fig_u_chiral}. Configurations at $t=14.7$ and 16.0~{ps} show that the radial breathing oscillation is disturbed gradually by the seventh and eighth modes (with three-fold symmetry) in Fig.~\ref{fig_u_chiral} due to the nonlinear effect relating to large actuation energy. The radial breathing oscillation is completely replaced by these three-fold modes after $t=18.6$~{ps}. These three-fold oscillation is clearly demonstrated by the two configurations at $t=21.0$ and 22.3~{ps}, which exactly follow the eigen vector of the two three-fold modes. This shows that the three-fold mode is also a type of stable oscillation, i.e this oscillation does not destroy the tubal structure. However, after $t=94.7$~{ps}, the fifth and sixth modes (with two-fold symmetry) in Fig.~\ref{fig_u_chiral} start to be excited. These two-fold modes are unstable due to their special vibration morphology and their low frequencies. A low frequency indicates that these modes are able to deform the tubal structure quite a lot with small amount of energy. Indeed, the tube starts to be squashed by these two-fold modes after $t=105.1$~{ps}.

\section{Conclusion}
In conclusion, we study the lattice dynamics properties of the SWMoS$_{2}$NT with arbitrary chiralities. In the construction of the $9\times 9$ dynamical matrix, the force constant matrix is calculated from the SW potential and the phonon modes are denoted by helical quantum numbers, which correspond to the line group in the system. The frequency and eigen vector of the low-frequency phonon modes are analyzed. The frequency of the RBM is found to be inversely proportional to the tube diameter as $\omega=665.3/d$~{cm$^{-1}$}, where $d$ is the tube diameter. We perform the molecular dynamics simulation to investigate the radial breathing mechanical oscillation, and find that the life time of the radial breathing oscillation decays exponentially with increasing oscillation energy. More specifically, in the initial stage, this radial breathing oscillation is disturbed by the three-fold modes. The SWMoS$_{2}$NTs are squashed by the two-fold modes in the final stage, if the oscillation is actuated into nonlinear regime.

\textbf{Acknowledgements} The work is supported by the Recruitment Program of Global Youth Experts of China (JWJ) and the German Research Foundation (DFG).

%
\end{document}